\documentclass[sigconf]{acmart}
\renewcommand\footnotetextcopyrightpermission[1]{} 
\usepackage{amsmath} 
\usepackage{amsfonts}

\usepackage{eqnarray}
\usepackage{amssymb}
\usepackage{amsthm}
\usepackage{natbib}
\usepackage{subfigure} 
\usepackage{paralist}
\usepackage{booktabs}
\usepackage{color}
\usepackage{microtype}
\usepackage{balance}

\DeclareGraphicsExtensions{.pdf,.jpeg,.png,.epbibdesks}
\usepackage[T1]{fontenc}
\usepackage{epstopdf}
\usepackage{graphicx}

\usepackage[algo2e]{algorithm2e}
\allowdisplaybreaks

\theoremstyle{remark}

\begin{document}

\title{Personalized Thread Recommendation \\ for MOOC Discussion Forums}

\author{Andrew S. Lan}
\affiliation{\institution{Princeton University}}
\email{andrew.lan@princeton.edu}

\author{Jonathan C. Spencer}
\affiliation{\institution{Princeton University}}
\email{j.spencer@princeton.edu}

\author{Ziqi Chen}
\affiliation{\institution{HKUST}}
\email{zchenas@connect.ust.hk}

\author{Christopher G. Brinton}
\affiliation{\institution{Zoomi Inc.}}
\email{christopher.brinton@zoomiinc.com}

\author{Mung Chiang}
\affiliation{\institution{Purdue University}}
\email{chiang@purdue.edu}



\begin{abstract}
Social learning, i.e., students learning from each other through social interactions, has the potential to significantly scale up instruction in online education. In many cases, such as in massive open online courses (MOOCs), social learning is facilitated through discussion forums hosted by course providers. 
In this paper, we propose a probabilistic model for the process of learners posting on such forums, using point processes.  Different from existing works, our method integrates topic modeling of the post text, timescale modeling of the decay in post activity over time, and learner topic interest modeling into a single model, and infers this information from user data. Our method also varies the excitation levels induced by posts according to the thread structure, to reflect typical notification settings in discussion forums.  We experimentally validate the proposed model on three real-world MOOC datasets, with the largest one containing up to 6,000 learners making 40,000 posts in 5,000 threads. Results show that our model excels at thread recommendation, achieving significant improvement over a number of baselines, thus showing promise of being able to direct learners to threads that they are interested in more efficiently.  Moreover, we demonstrate analytics that our model parameters can provide, such as the timescales of different topic categories in a course.

\end{abstract}

\maketitle

{

\section{Introduction}
Online discussion forums have gained substantial traction over the past decade, and are now a significant avenue of knowledge sharing on the Internet.  Attracting learners with diverse interests and backgrounds, some platforms (e.g., Stack Overflow, MathOverflow) target specific technical subjects, while others (e.g., Quora, Reddit) cover a wide range of topics from politics to entertainment.

More recently, discussion forums have become a significant component of online education, enabling students in online courses to learn socially as a supplement to their studying of the course content individually \cite{slnefficiency}; social interactions between learners have been seen to improve learning outcomes \cite{peter}. In particular, massive open online courses (MOOCs) often have tens of thousands of learners within single sessions, making the social interactions via these forums critical to scaling up instruction \cite{moocgenerative}. In addition to serving as a versatile complement to self-regulated learning \cite{getoor}, research has shown that learner participation on forums can be predictive of learning outcomes \cite{kenrose}.

In this paper, we ask: \textit{How can we model the activity of individual learners in MOOC discussion forums?}  Such a model, designed correctly, presents several opportunities to optimize the learning process, including personalized news feeds to help learners sort through forum content efficiently, and analytics on factors driving participation.

\vspace{-0.0cm}
\subsection{Prior work on discussion forums}

\paragraph{Generic online discussion sites.}
There is vast literature on analyzing user interactions in online social networks (e.g., on Facebook, Google+, and Twitter). Researchers have developed methods for tasks including link prediction \cite{leskoveclinkprediction,jordanlinkprediction}, tweet cascade analysis \cite{gt,jordanpois}, post topic analysis \cite{twittermodel}, and latent network structure estimation \cite{networkhawkes,ningxia}. These methods are not directly applicable to modeling MOOC discussion forums since MOOCs do not support an inherent social structure; learners cannot become ``friends'' or ``follow'' one another.

Generic online discussion forums (e.g., Stack Overflow, Quora) have also generated substantial research.  Researchers have developed methods for tasks including question-answer pair extraction \cite{questionanswerpairs}, topic dynamics analysis \cite{dynamictopic}, post structure analysis \cite{discussioncrf}, and user grouping \cite{forumgrouping}.  While these types of forums also lack explicit social structure, MOOC discussion forums exhibit several unique characteristics that need to be accounted for. 
First, topics in MOOC discussion forums are mostly centered around course content, assignments, and course logistics \cite{moocgenerative}, making them far more structured than generic forums; thus, topic modeling can be used to organize threads and predict future activity. 
Second, there are no sub-forums in MOOCs: learners all post in the same venue even though their interests in the course vary.  Modeling individual interest levels on each topic can thus assist learners in navigating through posts.

\paragraph{MOOC forums.} A few studies on MOOC discussion forums have emerged recently.  The works in \cite{forumseededlda,moocsentiment} extracted forum structure and post sentiment information by combining unsupervised topic models with sets of expert-specified course keywords. In this work, our objective is to model learners' forum behavior, which requires analyzing not only the content of posts but also individual learner interests and temporal dynamics of the posts.

In terms of learner modeling, the work in \cite{mooccommunity} employed Bayesian nonnegative matrix factorization to group learners into communities according to their posting behavior.  This work relies on topic labels of each discussion post, though, which are either not available or not reliable in most MOOC forums.  The work in \cite{slnefficiency} inferred learners' topic-specific seeking and disseminating tendencies on forums to quantify the efficiency of social learning networks.  However, this work relies on separate models for learners and topics, whereas we propose a unified model. The work in \cite{waset} couples social network analysis and association rule mining for thread recommendation; while their approach considers social interactions among learners, they ignore the content and timing of posts.

As for modeling temporal dynamics, the work in \cite{moocgenerative} proposed a method that classifies threads into different categories (e.g., small-talk, course-specific) and ranks thread relevance for learners over time. This model falls short of making recommendations, though, since it does not consider learners individually. The work in \cite{rose} employed matrix factorization for thread recommendation and studied the effect of window size, i.e., recommending only threads with posts in a recent time window.  However, this model uses temporal information only in post-processing, which limits the insights it offers.  The work in \cite{epfl} focuses on learner thread viewing rather than posting behavior, which is different from our study of social interactions since learners view threads independently.

The model proposed in \cite{mozer} is perhaps most similar to ours, as it uses point processes to analyze discussion forum posts and associates different timescales with different types of posts to reflect recurring user behavior.  With the task of predicting which Reddit sub-forum a user will post in next, the authors base their point processes model on self-excitations, as such behavior is mostly driven by a user's own posting history.  Our task, on the contrary, is to recommend threads to learners taking a particular online course: here, excitations induced by other learners (e.g., explicit replies) can significantly affect a learner's posting behavior. As a result, the model we develop incorporates mutual excitation.  Moreover, \cite{mozer} labels each post based on the Reddit sub-forum it belongs to; no such sub-forums exist in MOOCs. 

\subsection{Our model and contributions}
In this paper, we propose and experimentally validate a probabilistic model for learners posting on MOOC discussion forums.  Our main contributions are as follows.

First, through point processes, our model captures several important factors that influence a learner's decision to post.  In particular, it models the probability that a learner makes a post in a thread at a particular point in time based on four key factors: (i) the interest level of the learner on the topic of the thread, (ii) the timescale of the thread topic (which corresponds to how fast the excitation induced by new posts on the topic decay over time), (iii) the timing of the previous posts in the thread, and (iv) the nature of the previous posts regarding this learner (e.g., whether they explicitly reply to the learner).  Through evaluation on three real-world datasets---the largest having more than 6,000 learners making more than 40,000 posts in more than 5,000 threads---we show that our model significantly outperforms several baselines in terms of thread recommendation, thus showing promise of being able to direct learners to threads they are interested in.

Second, we derive a Gibbs sampling parameter inference algorithm for our model. While existing work has relied on thread labels to identify forum topics, such metadata is usually not available for MOOC forum threads.  As a result, we jointly analyze the post timestamp information and the text of the thread by coupling the point process model with a topic model, enabling us to learn the topics and other latent variables through a single procedure.  

Third, we demonstrate several types of analytics that our model parameters can provide, using our datasets as examples. These include: (i) identifying the timescales (measured as half-lives) of different topics, from which we find that course logistics-related topics have the longest-lasting excitations, (ii) showing that learners are much (20-30 times) more likely to post again in threads they have already posted in, and (iii) showing that learners receiving explicit replies in threads are much (300-500 times) more likely to post again in these threads to respond to these replies.  
}

{

\section{Point Processes Forum Model}
An online course discussion forum is generally comprised of a series of threads, with each thread containing a sequence of posts and comments on posts. Each post/comment contains a body of text, written by a particular learner at a particular point in time. A thread can further be associated with a topic, based on analysis of the text written in the thread.  Figure~\ref{fig:illustration} (top) shows an example of a thread in a MOOC consisting of eight posts and comments. Moving forward, the terminology ``posting in a thread'' will refer to a learner writing either a post or a comment.

We postulate that a learner's decision to post in a thread at a certain point in time is driven by four main factors: (i) the learner's interest in the thread's topic, (ii) the timescale of the thread's topic, (iii) the number and timing of previous posts in the thread, and (iv) the learner's prior activity in the thread (e.g., whether there are posts that explicitly reply to the learner).  
The first factor is consistent with the fact that MOOC forums generally have no sub-forums: in the presence of diverse threads, learners are most likely to post in those covering topics they are interested in.  
The second factor reflects the observation that different topics exhibit different patterns of temporal dynamics.
The third factor captures the common options for thread-ranking that online forums provide to users, e.g., by popularity or recency; learners are more likely to visit those at the top of these rankings.  
The fourth factor captures the common setup of notifications in discussion forums: learners are typically subscribed to threads automatically once they post in them, and notified of any new posts (especially those that explicitly reply to them) in these threads.  
%
To capture these dynamics, we model learners' posts in threads as events in temporal point processes \cite{hawkesbook}, which will be described next.

\paragraph{Point processes.}  
A point process, the discretization of a Poisson process, is characterized by a rate function $\lambda(t)$ that models the probability that an event will happen in an infinitesimal time window $\mathrm{d}t$ \cite{hawkesbook}. 
Formally, the rate function at time $t$ is given by
\begin{align} \label{eq:pprate}
\lambda(t) = \mathbb{P}\left(\text{event in } [t, t+\mathrm{d}t)\right) = \lim_{\mathrm{d}t \rightarrow 0} \frac{N(t+\mathrm{d}t) - N(t)}{\mathrm{d}t},
\end{align} 
where $N(t)$ denotes the number of events up to time $t$ \cite{hawkesbook}. Assuming the time period of interest is $[0, T)$, the likelihood of a series of events at times $t_1, \ldots, t_N < T$ is given by:
\begin{align} \label{eq:lik}
\mathcal{L}(\{{t_i}\}_{i=1}^N) = \left( \prod_{i=1}^N \lambda(t_i)\right)  e^{-\int_0^T \lambda(\tau) \mathrm{d} \tau}. 
\end{align}
In this paper, we are interested in rate functions that are affected by excitations of past events (e.g., forum posts in the same thread).  Thus, we resort to Hawkes processes \cite{mozer}, which characterize the rate function at time $t$ given a series of past events at $t_1, \ldots, t_{N'} < t$ as 
\begin{align*} 
\lambda(t) = \mu + a  \sum_{i=1}^{N'} \kappa(t-t_i),
\end{align*}
where $\mu \geq 0$ denotes the constant background rate, $a \geq 0$ denotes the amount of excitation each event induces, i.e., the increase in the rate function after an event,\footnote{$a$ is sometimes referred to in literature as the impulse response \cite{networkhawkes}.} and $\kappa(\cdot): \mathbb{R}_+ \rightarrow [0,1]$ denotes a non-increasing decay kernel that controls the decay in the excitation of past events over time. In this paper, we use the standard exponential decay kernel $\kappa(t) = e^{-\gamma t}$, where $\gamma$ denotes the decay rate. Through our model, different decay rates can be associated with different topics \cite{mozer}; as we will see, this model choice enables us to categorize posts into groups (e.g., course content-related, small talk, or course logistics) based on their timescales, which leads to better model analytics.

\begin{figure}[t]
\centering
\includegraphics[width=1.0\columnwidth]{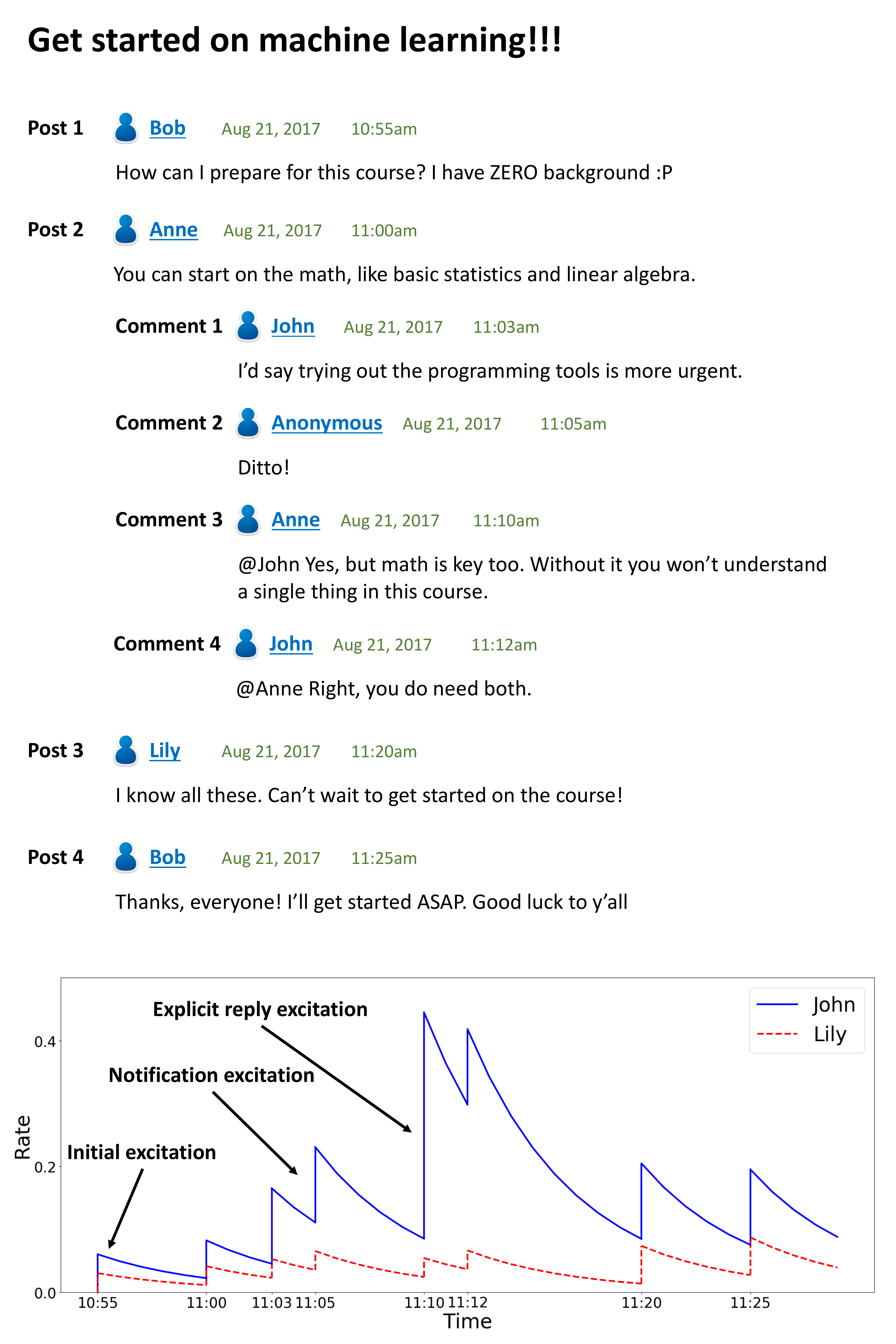}
\vspace{-0.0cm}
\caption{An example of how threads are structured in MOOC discussion forums (top) and an illustration of corresponding rate functions (bottom) for two learners in this thread. Different posts induce different amounts of excitation depending on whether and how they refer to the learner.}
\vspace{-0.0cm}
\label{fig:illustration}
\end{figure}

\paragraph{Rate function for new posts.} 

Let $U$, $K$, and $R$ denote the number of learners, topics, and threads in a discussion forum, indexed by $u$, $k$, and $r$, respectively.  We assume that each thread~$r$ functions independently, and that each learner's activities in each thread and on each topic are independent.  Further, let $z_r$ denote the topic of thread~$r$, and let $P_r$ denote the total number of posts in the thread, indexed by $p$; for each post~$p$, we use $u_p^r$ and $t_p^r$ to denote the learner index and time of the post, and we use $p^r_i(u)$ to denote the $i^\text{th}$ post of learner~$u$ in thread~$r$.  Note that posts in a thread are indexed in chronological order, i.e., $p < p'$ if and only if $t_p^r < t_{p'}^r$.  Finally, let $\gamma_k \geq 0$ denote the decay rate of each topic and let $a_{u,k}$ denote the interest level of learner~$u$ on topic~$k$.  We model the rate function that characterizes learner~$u$ posting in thread $r$ (on topic $z_r = k$) at time $t$ given all previous posts in the thread (i.e., posts with $t_p^r < t$) as
\begin{align} \label{eq:hawkesrate}
\lambda_{u,k}^r (t)  =  \left \{ \!\! \begin{array}{ll}
a_{u,k} \sum_p e^{-\gamma_k (t - t_p^r)}  & \!\!\!\!\!\!\!\!\!\!\!\! \text{if} \;\; t < t_{p_1^r (u)}^r \\[0.2cm]
a_{u,k} \sum_{p: p < p_1^r (u)} e^{-\gamma_k (t - t_p^r)} \\[0.1cm]
\, + \alpha \, a_{u,k} \sum_{p: p \geq p_1^r (u), u \notin d_p^r} e^{-\gamma_k (t - t_p^r)} \\[0.1cm]
\, + \beta \alpha \, a_{u,k} \sum_{p: u \in d_p^r} e^{-\gamma_k (t - t_p^r)} & \!\!\!\!\!\!\!\!\!\!\!\! \text{if} \;\; t \geq t_{p_1^r (u)}^r.
\end{array} \right.
\end{align}

In our model, $a_{u,k}$ characterizes the base level of excitation that learner~$u$ receives from posts in threads on topic~$k$, which captures the different interest levels of learners on different topics.  The exponential decay kernel models a topic-specific decay in excitation of rate $\gamma_k$ from the time of the post.

Before $t_{p_1^r(u)}^r$ (the timestamp of the first post learner~$u$ makes in thread~$r$), learner~$u$'s rate is given solely by the number and recency of posts in $r$ ($t_{p_1^r (u)}^r = \infty$ if the learner never posts in this thread), while all posts occurring after $t_{p_1^r(u)}^r$ induce additional excitation characterized by the scalar variable $\alpha$.  
This model choice captures the common setup in MOOC forums that learners are automatically subscribed to threads after they post in them. Therefore, we postulate that $\alpha > 1$, since new post notifications that come with thread subscriptions tend to increase a learner's chance of viewing these new posts, in turn increasing their likelihood of posting again in these threads.  The observation of users posting immediately after receiving notifications is sometimes referred to as the ``bursty'' nature of posts on social media \cite{gt}.  

We further separate posts made after $t_{p_1^r(u)}^r$ by whether or not they constitute \textit{explicit replies} to learner~$u$.   
A post $p'$ is considered to be an explicit reply to a post $p$ in the same thread $r$ if $t^r_{p'} > t^r_p$ and one of the following conditions is met: (i) $p'$ makes direct reference (e.g., through name or the @ symbol) to the learner who made post $p$, or (ii) $p'$ is the first comment under $p$.\footnote{In this work, we restrict ourselves to these two concrete types of explicit replies; analyzing other, more ambiguous types is left for future work.}  $d_p^r$ in (\ref{eq:hawkesrate}) denotes the set of explicit recipients of $p$, i.e., if $p$ is an explicit reply to learner $u$, then $u \in d_p^r$, while if $p$ is not an explicit reply to any learners then $d_p^r = \emptyset$.  
This setup captures the common case of learners being notified of posts that explicitly reply to them in a thread.  The scalar $\beta$ characterizes the additional excitation these replies induce; we postulate that $\beta > 1$, i.e., the personal nature of explicit replies to learners' posts tends to further increase the likelihood of them posting again in the thread (e.g., to address these explicit replies).

\paragraph{Rate function for initial posts.} 
We must also model the process of generating the initial posts in threads. We characterize the rate function of these posts as time-invariant:
\begin{align} \label{eq:seekrate}
\lambda_{u,k}^r (t) = \mu_{u,k},
\end{align}
where $\mu_{u,k}$ denotes the background posting rate of learner~$u$ on topic~$k$.  Separating the initial posts in threads from future posts in this way enables us to model learners' knowledge seeking (i.e., starting threads) and knowledge disseminating (i.e., posting responses in threads) behavior \cite{slnefficiency}, through the background ($\mu_{u,k}$) and excitation levels ($a_{u,k}$), respectively.
\paragraph{Post text modeling.} 
Finally, we must also model the text of each thread.  Given the topic $z_r = k$ of thread~$r$, we model $\mathcal{W}_r$---the bag-of-words representation of the text in $r$ across all posts---as being generated from the standard latent Dirichlet allocation (LDA) model \cite{bleilda}, with topic-word distributions parameterized by $\boldsymbol{\phi}_k$.  Details on the LDA model and the posterior inference step for $\boldsymbol{\phi}_k$ via collapsed Gibbs sampling in our parameter inference algorithm are omitted for simplicity of exposition.
\paragraph{Model intuition.} 
Intuitively, a learner will browse existing threads in the discussion forum when they are interested in a particular topic.  If a relevant thread exists, they may make their first post there (e.g., Comment~1 by John under Post~2, in Figure~\ref{fig:illustration}), with the rate at which this occurs being governed by the previous activity in the thread (posts at times $t_p^r < t$) and the learner's interest level in the topic of the thread ($a_{u,k}$).  Together with the exponential decay kernel, this model setting reflects the observation that discussion forum threads are often sorted by recency (the time of last post) and popularity (typically quantified by the number of replies).  Additionally or alternatively, if no such thread exists, the learner may decide to start a new thread on the topic (e.g., Post~1 by Bob), depending on their background rate ($\mu_{u,k}$).  Once the learner has posted in a thread, they will receive notifications of new posts there (e.g., Lily will be notified of Post~4), which induces higher levels of excitation ($\alpha > 1$); the personal nature of explicit replies to their posts (e.g., Anne's mention of John in Comment~3 under Post~2) will induce even higher levels of excitation ($\beta > 1$).

}

{

\section{Parameter Inference} 
We now derive the parameter inference algorithm for our model.  
We perform inference using Gibbs sampling, i.e., iteratively sampling from the posterior distributions of each latent variable, conditioned on the other latent variables. The detailed steps are as follows:
\begin{itemize}
\item[1. Sample $z_r$.] To sample from the posterior distribution of the topic of each thread, $z_r$, we put a uniform prior over each topic and arrive at the posterior
\begin{align*}
P(z_r = k\! \mid \!\ldots) \propto & \; P(\mathcal{W}_r\! \mid \!z_r) \\ 
& \quad \cdot \prod_{k'} P(\{t_1^{r'}\}_{r': z_{r'} = k', u^r_1 = u^{r'}_1}\! \mid \!\mu_{u_1^r,k'}) \\
& \quad \cdot  \prod_u P(\{t_p^r\}_{p: u_p^r = u}\! \mid \!a_{u,k}, \alpha, \beta, \gamma_k),
\end{align*}
\sloppy
where $\ldots$ denotes all variables except $z_r$. $P(\mathcal{W}_r\!\! \mid \!\!z_r)$ denotes the likelihood of observing the text of thread $r$ given its topic. $P(\{t_1^{r'}\}_{r': z_{r'} = k', u^r_1 = u^{r'}_1}\!\! \mid \!\mu_{u_1^r,k'})$ denotes the likelihood of observing the sequence of initial thread posts on topic~$k'$ made by the learner who also made the initial post in thread $r$;\footnote{If $\mu_1^r$ is not the initial poster in any thread $r'$ with $z_{r'} = k'$, then $\{t^{r'}_1\} = \emptyset$.} this is given by substituting (\ref{eq:seekrate}) into (\ref{eq:lik}) as
\begin{align} \label{eq:mulik}
\notag & P(\{t_1^{r'}\}_{r': z_{r'} = k', u^r_1 = u^{r'}_1}\! \mid \!\mu_{u_1^r,k'}) \\
& \qquad\qquad = \mu_{u_1^r,k'}^{\sum_{r'} \mathbf{1}_{u_1^r = u^{r'}_1, z_{r'} = k'}} \cdot e^{-\mu_{u_1^r,k'}T}, 
\end{align}
where $\mathbf{1}_x$ denotes the indicator function that takes the value $1$ when condition $x$ holds and $0$ otherwise.
$P(\{t_p^r\}_{p: u_p^r = u}\! \mid \!a_{u,k}, \alpha, \beta, \gamma_k)$ denotes the likelihood of observing the sequence of posts made by learner~$u$ in thread $r$,\footnote{If $u$ has not posted in $r$, then $\{t_p^r\} = \emptyset$.} 
given by 
\begin{align} \label{eq:alik}
\notag & P(\{t_p^r\}_{p: u_p^r = u}\! \mid \!a_{u,k}, \alpha, \beta, \gamma_k) \\ 
& \qquad = \left(\prod_{p: u^r_p = u} \lambda_{u,z_r}^r(t^r_p)\right) \left( e^{-\int_0^T \lambda_{u,z_r}^r (t) \mathrm{d}t} \right),
\end{align}
where the rate function $\lambda_{u,k}^r (t)$ for learner $u$ in thread $r$ (with topic $k$) is given by (\ref{eq:hawkesrate}). 
\item[2. Sample $\gamma_k$.] There is no conjugate prior distribution for the excitation decay rate variable $\gamma_k$.  Therefore, we resort to a pre-defined set of decay rates $\gamma_k \in \{\gamma_s\}_{s=1}^S$.  We put a uniform prior on $\gamma_k$ over values in this set, and arrive at the posterior given by
\begin{align*}
& P(\gamma_k = \gamma_s\! \mid \!\ldots) \propto  \prod_{r: z_r = k}  \prod_u P(\{t_p^r\}_{p: u_p^r = u}\! \mid \!a_{u,k}, \alpha, \beta, \gamma_s).
\end{align*}
\item[3. Sample $\mu_{u,k}$.] The conjugate prior of the learner background topic interest level variable $\mu_{u,k}$ is the Gamma distribution.  Therefore, we put a prior on $\mu_{u,k}$ as $\mu_{u,k} \sim \mbox{Gam}(\alpha_\mu,\beta_\mu)$ and arrive at the posterior distribution
$$P(\mu_{u,k}\! \mid \!\ldots) \propto \mbox{Gam}(\alpha_\mu',\beta_\mu')$$
where
\begin{align*}
\alpha_\mu' = \alpha_\mu +  \sum_r \mathbf{1}_{u_1^r = u, z_r = k}, \qquad \beta_\mu' = \beta_\mu + T.
\end{align*}
\item[4. Sample $a_{u,k}$, $\alpha$, and $\beta$.] The latent variables $\alpha$ and $\beta$ have no conjugate priors. As a result, we introduce an auxiliary latent variable \cite{networkhawkes,jordanpois} $e_p^r$ for each post $p$, where $e_{p'}^r = p$ means that post~$p$ is the ``parent'' of post~$p'$ in thread~$r$, i.e., post~$p'$ was caused by the excitation that the previous post~$p$ induced.  We first sample the parent variable for each post $p$ according to
\begin{align*}
P(e_{p'}^r = p)  \propto  a^r(p,p')  e^{- \gamma_{z_r} (t_{p'}^r - t_p^r)},
\end{align*}
where $a^r(p,p') \in \{a_{u_{p'}^r,z_r}, \alpha a_{u_{p'}^r,z_r}, \beta \alpha a_{u_{p'}^r,z_r} \}$ depending on the relationship between posts~$p$ and $p'$ from our model, i.e., whether $p'$ is the first post of $u_{p'}$ in the thread, and if not, whether $p$ is an explicit reply to $u_{p'}$. In general, the set of possible parents of $p$ is all prior posts $1, \ldots, p-1$ in $r$, but in practice, we make use of the structure of each thread to narrow down the set of possible parents for some posts.\footnote{For example, in Fig.~\ref{fig:illustration}, Post~2 is the only possible parent post of Comment~1 below, as Comment~1 is an explicit reply to Post~2. We omit the details of this step for simplicity of exposition.}

With these parent variables, we can write $\mathcal{L}(\{t_p^r\}_{p: u_p^r=u})$, the likelihood of the series of posts learner~$u$ makes in thread~$r$ as
\begin{align*}
& \mathcal{L} =  \prod_r \mathcal{L}(\{t_p^r\}_{p=1}^{P_r}) =  \prod_r  \prod_u \mathcal{L}(\{t_p^r\}_{p: u_p^r=u}),
\end{align*}
where $\mathcal{L}(\{t_p^r\}_{p: u_p^r=u})$ denotes the likelihood of the series of posts learner~$u$ makes in thread~$r$. We can then expand the likelihood using the parent variables as
\begin{align*}
& \mathcal{L}(\{t_p^r\}_{u_p^r=u}) =   \prod_{p: p < p_1^r(u)} e^{-\frac{a_{u,z_r}}{\gamma_{z_r}}(1-e^{-\gamma_{z_r}(T - t_p^r)})} \\
& \quad \left(   \prod_{p': u_{p'}^r = u, e_{p'}^r = p} a_{u,z_r} e^{-\gamma_{z_r}(t_{p'}^r - t_p^r)} \right) \\
& \quad \quad \quad\quad\quad \cdot  \prod_{p: p \geq p_1^r (u), u \notin d_p^r} e^{-\frac{\alpha a_{u,z_r}}{\gamma_{z_r}}(1-e^{-\gamma_{z_r}(T - t_p^r)})} \\
& \quad \left( \prod_{p': u_{p'}^r = u, e_{p'}^r = p} \alpha a_{u,z_r} e^{-\gamma_{z_r}(t_{p'}^r - t_p^r)} \right) \\
& \quad \quad \quad\quad\quad \cdot  \prod_{p: u \in d_p^r} e^{-\frac{\beta \alpha a_{u,z_r}}{\gamma_{z_r}}(1-e^{-\gamma_{z_r}(T - t_p^r)})} \\
& \quad \left(  \prod_{p': u_{p'}^r = u, e_{p'}^r = p} \beta \alpha a_{u,z_r} e^{-\gamma_{z_r}(t_{p'}^r - t_p^r)} \right). 
\end{align*}

We now see that Gamma distributions are conjugate priors for $a_{u,k}$, $\alpha$, and $\beta$. Specifically, if $a_{u,k} \sim \mbox{Gam}(\alpha_a,\beta_a)$, its posterior is given by $P(a_{u,k} | \ldots) \sim \mbox{Gam}(\alpha_a',\beta_a')$ where
\begin{align*}
& \alpha_a' = \alpha_a +  \sum_{r:z_r = k}  \sum_p \mathbf{1}_{u_{p}^r = u}, \\
& \beta_a' = \beta_a +  \sum_{r:z_r = k} \Big(\sum_{p: p < p_1^r(u)} \frac{1}{\gamma_k} (1-e^{-\gamma_k (T - t_p^r)}) \\
& \quad +  \sum_{p:p \geq p_1^r(u), u \notin d_p^r} \frac{\alpha}{\gamma_k} (1-e^{-\gamma_k (T - t_p^r)}) \\
& \quad +  \sum_{p: u \in d_p^r} \frac{\beta \alpha}{\gamma_k} (1-e^{-\gamma_k (T - t_p^r)}) \Big).
\end{align*}

\sloppy
Similarly, if $\alpha \sim \mbox{Gam}(\alpha_\alpha,\beta_\alpha)$, the posterior is $P(\alpha | \ldots) \sim \mbox{Gam}(\alpha_\alpha',\beta_\alpha')$ where
\begin{align*}
& \alpha_\alpha' = \alpha_\alpha +  \sum_r  \sum_p  \sum_{p'} \mathbf{1}_{e_{p'}^r = p, p \geq p_1^r (u_{p'}^r)}, \\
& \beta_\alpha' = \beta_\alpha +  \sum_r  \sum_u \Big(  \sum_{p:p \geq p_1^r(u), u \notin d_p^r} \frac{a_{u,z_r}}{\gamma_{z_r}}  (1-e^{-\gamma_{z_r} (T - t_p^r)}) \\ & \quad +  \sum_{p: u \in d_p^r} \frac{\beta a_{u,z_r}}{\gamma_{z_r}} (1-e^{-\gamma_{z_r} (T - t_p^r)}) \Big).
\end{align*}

Finally, if $\beta \sim \mbox{Gam}(\alpha_\beta,\beta_\beta)$, the posterior is $P(\beta | \ldots) \sim \mbox{Gam}(\alpha_\beta',\beta_\beta')$ where
\begin{align*}
& \alpha_\beta' = \alpha_\beta +  \sum_r  \sum_p \sum_{p'} \mathbf{1}_{e_{p'}^r = p, u_{p'}^r \in d_p^r}, \\
& \beta_\beta' = \beta_\beta +  \sum_r  \sum_u  \sum_{p: u \in d_p^r} \frac{\alpha a_{u,z_r}}{\gamma_{z_r}} (1-e^{-\gamma_{z_r} (T - t_p^r)}).
\end{align*}

\end{itemize}

We iterate the sampling steps 1--4 above after randomly initializing the latent variables according to their prior distributions.  After a burn-in period, we take samples from the posterior distribution of each variable over multiple iterations, and use the average of these samples as its estimate. 
}

{

\section{Experiments}
In this section, we experimentally validate our proposed model using three real-world MOOC discussion forum datasets. In particular, we first show that our model obtains substantial gains in thread recommendation performance over several baselines. Subsequently, we demonstrate the analytics on forum content and learner behavior that our model offers.


\begin{figure*}[t]
\centering
\subfigure[\texttt{ml}]{\includegraphics[width=0.69\columnwidth]{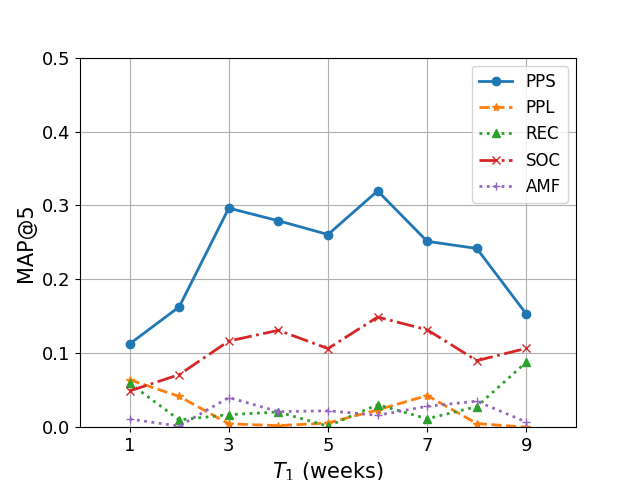}}
\subfigure[\texttt{algo}]{\includegraphics[width=0.69\columnwidth]{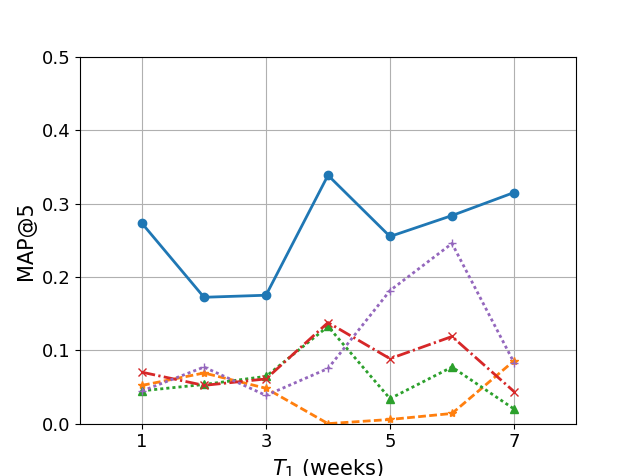}}
\subfigure[\texttt{comp}]{\includegraphics[width=0.69\columnwidth]{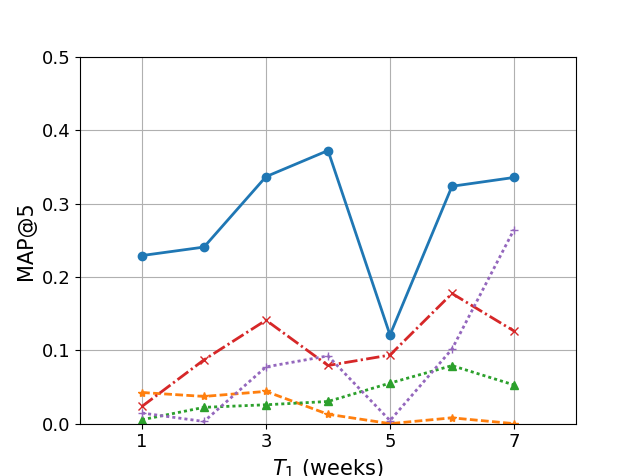}}
\vspace{-0.0cm}
\caption{Plot of recommendation performance over different lengths of the training time window $T_1$ on all datasets. Our model significantly outperforms every baseline.}
\label{fig:sweeptr}
\end{figure*}

\vspace{-0.2cm}
\subsection{Datasets}
We obtained three discussion forum datasets from 2012 offerings of MOOCs on Coursera: Machine Learning (\texttt{ml}), Algorithms, Part I (\texttt{algo}), and English Composition I (\texttt{comp}).  The number of threads, posts and learners appearing in the forums, and the duration (the number of weeks with non-zero discussion forum activity) of the courses are given in Table~\ref{tab:stats}.

\begin{table}
\scalebox{1.0}{
\begin{tabular}{ccccc}
Dataset & Threads & Posts & Learners & Weeks \\ \toprule
{\tt ml} & 5,310 & 40,050 & 6,604 & 15 \\
{\tt algo} & 1,323 & 9,274 & 1,833 & 9 \\
{\tt comp} & 4,860 & 17,562 & 3,060 & 14 \\
\end{tabular}}
\vspace{0.2cm}
\caption{Basic statistics on the datasets.}
\label{tab:stats}

\end{table}

Prior to experimentation, we perform a series of pre-processing steps. 
First, we prepare the text for topic modeling by (i) removing non-ascii characters, url links, punctuations and words that contain digits, (ii) converting nouns and verbs to base forms, (iii) removing stopwords,\footnote{We use the stopword list in the Python natural language toolkit (\url{http://www.nltk.org/}) that covers 15 languages.} and (iv) removing words that appear fewer than 10 times or in more than 10\% of threads. 
Second, we extract the following information for each post: (i) the ID of the learner who made the post ($u_p^r$), (ii) the timestamp of the post ($t_p^r$), and (iii) the set of learners it explicitly replies to as defined in the model ($d_p^r$). 
For posts made anonymously, we do not include rates for them ($\lambda_{u,k}^r(t)$) when computing the likelihood of a thread, but we do include them as sources of excitation for non-anonymous learners in the thread.

\begin{figure}[t]
\centering
\includegraphics[width=0.75\columnwidth]{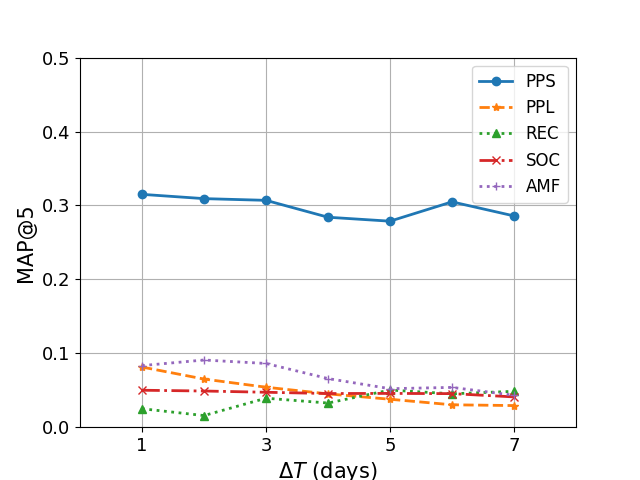}
\vspace{-0.0cm}
\caption{Recommendation performance of the algorithms for varying testing window length $\Delta T$ on the \texttt{algo} dataset. The point process-based algorithms have highest performance and are more robust to $\Delta T$.}
\label{fig:sweepte}
\end{figure}

\subsection{Thread recommendation}

\paragraph{Experimental setup.}
We now test the performance of our model on personalized thread recommendation.  We run three different experiments, splitting the dataset based on the time of each post. The training set includes only threads initiated during the time interval $[0, T_1)$, i.e., $\{r: t_1^r \in [0, T_1)\}$, and only posts on those threads made before $T_1$, i.e., $\{p: t_p^r \leq T_1\}$. The test set contains posts made in time interval $[T_1, T_2)$, i.e., $\{p: t_p^r \in [T_1, T_2)\}$, but excludes new threads initiated during the test interval.

In the first experiment, we hold the length of the testing interval fixed to 1 day, i.e., $\Delta T = T_2 - T_1 = 1\,\text{day}$, and vary the length of the training interval as $T_1 \in \{1\,\text{week}, \ldots, W-1\,\text{weeks} \}$, where $W$ denotes the number of weeks that the discussion forum stays active.  We set $W$ to 10, 8, and 8 for ml, comp, and algo, respectively, to ensure the number of posts in the testing set is large enough.  These numbers are less than those in Table~\ref{tab:stats} since learners drop out during the course, which leads to decreasing forum activity. 
In the second experiment, we hold the length of the training interval fixed at $W-1$ weeks and vary the length of the testing interval as $\Delta T \in \{ 1\,\text{day}, \ldots, 7\, \text{days}\}$. 
In the first two experiments, we fix $K = 5$, while in the third experiment, we fix the length of the training and testing intervals to $7$ weeks and $1$ week, respectively, and vary the number of latent topics as $K \in \{2, 3, \ldots, 10, 12, 15, 20\}$.   

For training, we set the values of the hyperparameters to $\alpha_a = \alpha_\mu = 10^{-4}$, and $\beta_a = \beta_\mu = \alpha_\alpha = \beta_\alpha = \alpha_\beta = \beta_\beta = 1$.  We set the pre-defined decay rates $\{\gamma_s\}_{s=1}^S$ to correspond to half-lives (i.e., the time for the excitation of a post to decay to half of its original value) ranging from minutes to weeks.  We run the inference algorithm for a total of $200$ iterations, with $100$ of these being burn-in iterations.\footnote{We observe that the Markov chain achieves reasonable mixing after about $50$ iterations.}

\paragraph{Baselines.}  We compare the performance of our point process model (PPS) against four baselines: (i) Popularity (PPL), which ranks threads from most to least popular based on the total number of posts in each thread during the training time interval; (ii) Recency (REC), which ranks threads from newest to oldest based on the timestamp of their most recent post; (iii) Social influence (SOC), a variant of our PPS model that replaces learner topic interest levels with learner social influences (the ``Hwk'' baseline in \cite{gt}); and (iv) Adaptive matrix factorization (AMF), our implementation of the matrix factorization-based algorithm proposed in \cite{rose}.

To rank threads in our model for each learner, we calculate the probability that learner~$u$ will reply to thread~$r$ during the testing time interval as
\begin{align*}
P(u\mathrm{~posts~in~}r) & =  \sum_k P(u\mathrm{~posts~in~}r\! \mid \!z_r=k) \,P(z_r=k) \nonumber \\
&=  \sum_k \Big(1- e^{-\int_{T_1}^{T_2}\lambda_{u,k}^r(t)\mathrm{d}t} \Big)\, P(z_r=k).  
\end{align*}
The rate function $\lambda_{u,k}^r(t)$ is given by (\ref{eq:hawkesrate}). $P(z_r=k)$ is given by
\begin{align*}
P(z_r=k) &\propto P(z_r=k\! \mid \!u_1^r) \, P(\mathcal{W}_r\! \mid \!z_r=k) \\ 
& \qquad \cdot  \prod_u P(\{t_p^r\}_{p: u_p^r = u, t_p^r < T_1}\! \mid \!z_r=k),
\end{align*}
where the likelihoods of the initial post and other posts are given by (\ref{eq:lik}) and (\ref{eq:mulik}), and the thread text likelihood $P(\mathcal{W}_r\!\! \mid \!\!z_r=k)$ is given by the standard LDA model. The threads are then ranked from highest to lowest posting probability.

\vspace{-0.0cm}
\paragraph{Evaluation metric.}
We evaluate recommendation performance using the standard mean average precision for top-N recommendation (MAP@N) metric. This metric is defined by taking the mean (over all learners who posted during the testing time interval) of the average precision
\begin{align*}
AP_u \text{@}N = \sum_{n=1}^{N} \frac{P_u \text{@} n \cdot \mathbf{1}_{\text{u posted in thread}\,r_u(n)}}{\min\{|\mathcal{R}_u|,N\}},
\end{align*}
where $\mathcal{R}_u$ denotes the set of threads learner~$u$ posted in during the testing time interval $[T_1, T_2)$, $r_u(n)$ denotes the $n^\text{th}$ thread recommended to the learner, $P_u \text{@} n$ denotes the precision at $n$, i.e., the fraction of threads among the top $n$ recommendations that the learner actually posted in. We use $N = 5$ in the first two experiments, and vary $N \in \{ 3, 5, 10\}$ in the third experiment.

\begin{figure}[t]
\centering
\includegraphics[width=0.75\columnwidth]{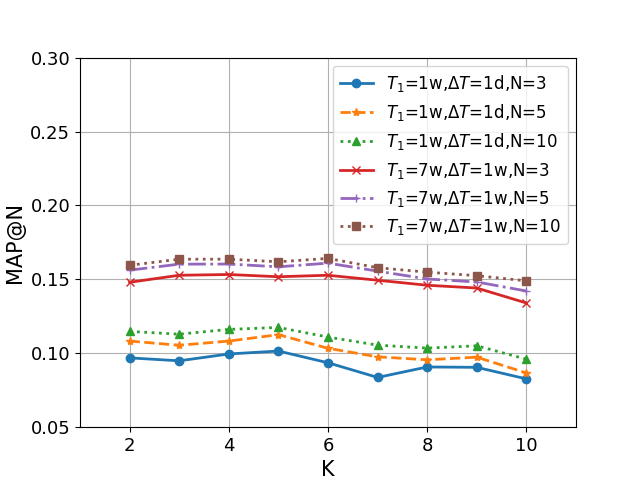}
\vspace{-0.0cm}
\caption{Plot of recommendation performance of our model over the number of topics $K$ on the \texttt{ml} dataset. The best performance is obtained at $K \approx 5$, though performance is stable for $K \leq 10$.}
\label{fig:sweepk}
\end{figure}

\begin{table*}[tp]
\centering
\scalebox{1.0}{
\begin{tabular}{lll}
\toprule
Topic & Half-life & Top words \\
\midrule
1 & 4 hours & gradient, row, element, iteration, return, transpose, logistic, multiply, initial, regularization \\
\midrule
2 &  4 hours & layer, classification, probability, neuron, unit, hidden, digit, nn, sigmoid, weight \\
\midrule
3 & 1 day & interest, group, computer, Coursera, study, hello, everyone, student, learning, software \\
\midrule
4 & 1 day & Coursera, deadline, professor, hard, score, certificate, review, experience, forum, material \\
\midrule
5 & 1 week & screenshot, speed, player, subtitle, chrome, firefox, summary, reproduce, open, graph \\
\bottomrule
\end{tabular}
}
\vspace{3mm}
\caption{Estimated half-lives and highest constituent words (obtained by sorting the estimated topic-word distribution parameter vectors $\phi_k$) for selected topics in the {\tt ml} dataset with at least 100 threads. Different types of topics (course content-related, small-talk, or course logistics) exhibit different half-lives.} \label{tbl:topics}
\end{table*}

\paragraph{Results and discussion.}
Fig.~\ref{fig:sweeptr} plots the recommendation performance of our model and the baselines over different lengths of the training time window $T_1$ for each dataset. Overall, we see that our model significantly outperforms the baselines in each case, achieving 15\%-400\% improvement over the strongest baseline.\footnote{Note that these findings are consistent across each dataset. Moving forward, we present one dataset in each experiment unless differences are noteworthy.}  The fact that PPS outperforms the SOC baseline confirms our hypothesis that in MOOC forums, learner topic preference is a stronger driver of posting behavior than social influence, consistent with the fact that most forums do not have an explicit social network (e.g., of friends or followers). 
The fact that PPS outperforms the AMF baseline emphasizes the benefit of the temporal element of point processes in capturing the dynamics in thread activities over time, compared to the (mostly) static matrix factorization-based algorithms.  
Note also that as the amount of training data increases in the first several weeks, the recommendation performance tends to increase for the point processes-based algorithms while decreasing for PPL and REC.  The observed fluctuations can be explained by the decreasing numbers of learners in the test sets as courses progress, since they tend to drop out before the end (see also Fig.~\ref{fig:posttimes}).

Fig.~\ref{fig:sweepte} plots the recommendation performance over different lengths of the testing time window $\Delta T$ for the \texttt{algo} dataset. As in Fig.~\ref{fig:sweeptr}, our model significantly outperforms every baseline.  
We also see that recommendation performance tends to decrease as the length of the testing time window increases, but while the performance of point process-based algorithms decay only slightly, the performance of the PPL and AMF baselines decrease significantly (by around 50\%). This observation suggests that our model excels at modeling long-term learner posting behavior.

Finally, Fig.~\ref{fig:sweepk} plots the recommendation performance of the PPS model over different numbers of topics $K$ for the \texttt{ml} dataset, for different choices of $N$, $T_1$ and $\Delta T$.  In each case, the performance rises slightly up to $K \approx 5$ and then drops for larger values (when overfitting occurs).  Overall, the performance is relatively robust to $K$, for $K \leq 10$. 

\begin{figure}[t]
\centering
\includegraphics[width=0.75\columnwidth]{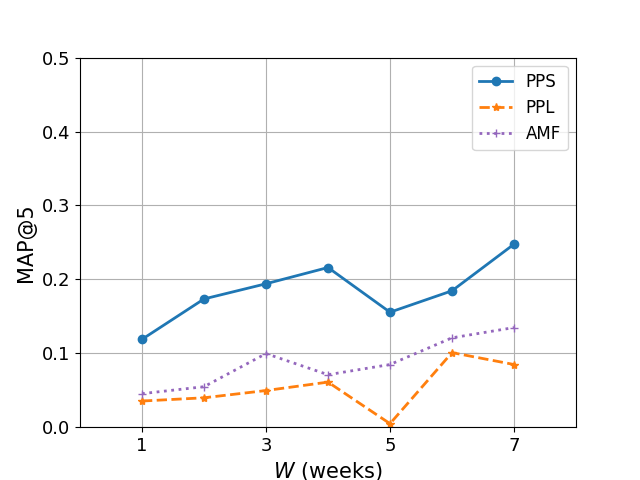}
\vspace{-0.0cm}
\caption{Direct comparison of our model against the AMF and PPL baselines using the experimental setup in \cite{rose} on the \texttt{comp} dataset. Our model again significantly outperforms both baselines.}
\label{fig:frose}
\end{figure}

\subsection{Direct comparison with AMF}

The MAP@5 values we obtained for both the AMF and PPL baselines are significantly less than those reported in \cite{rose}, where AMF is proposed.  To investigate this, we also perform a direct, head-to-head comparison between our model and these baselines under our closest possible replication of the experimental setting in \cite{rose}.  In particular, we train on threads that have non-zero activity between weeks $W-1$ and $W$, fix the testing time window to $\Delta T = \text{1 week}$, and set $K = 6$.  Since the exact procedures used in \cite{rose} to select the latent dimension in the ``content level model,'' to select the number of close peers in the ``social peer connections'', and to aggregate these two into a single model for matrix factorization in AMF are not clear, we sweep over a range of values for these parameters and choose the values that maximize the performance of AMF.

Fig.~\ref{fig:frose} compares the MAP@5 performance of our model against that of the PPL and AMF baselines for a range of values of $W$ on the \texttt{comp} dataset (as in previous experiments, results on the other two datasets are similar).  We see again that our model significantly outperforms both AMF and PPL in each case.  Moreover, while AMF consistently outperforms PPL in agreement with the results in \cite{rose}, the MAP@5 values of both baselines are significantly less than the values of $0.3$ reported in \cite{rose}.  We also emphasize that setting the length of the testing window to 1 week is too coarse of a timescale for thread recommendation in the MOOC discussion forum setting, where new discussions may emerge on a daily basis due to the release of new learning content, homework assignments, or exams.

\subsection{Model analytics}
Beyond thread recommendation, we also explore a few types of analytics that our trained model parameters can provide. For this experiment, we set $K = 10$ in order to achieve finer granularity in the topics; we found that this leads to more useful analytics.


\begin{figure}[tp]
\centering
\includegraphics[width=0.75\columnwidth]{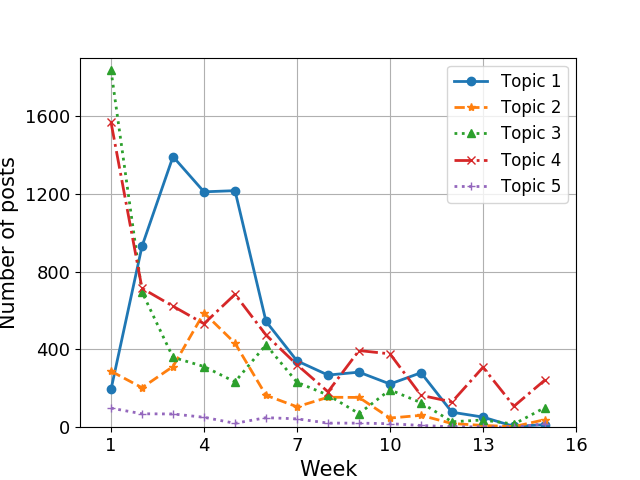}
\vspace{-0.0cm}
\caption{Plot of the total number of posts on each topic week-by-week in the \texttt{ml} dataset.  The week-to-week activity levels vary significantly across topics.}
\vspace{-0.0cm}
\label{fig:posttimes}
\end{figure}

\begin{table}[tp]
\centering
\scalebox{1.0}{
\begin{tabular}{cccc}
Dataset & \texttt{ml} & \texttt{algo} & \texttt{comp} \\ \toprule
$\widehat{\alpha}$ & 29.0 & 23.3 & 33.6 \\
$\widehat{\beta}$ & 19.2 & 12.2 & 10.6
\end{tabular}}
\vspace{0.2cm}
\caption{Estimated levels of additional excitation brought by new activity notifications and explicit replies.} \label{tbl:alphabeta}
\end{table}

\paragraph{Topic timescales and thread categories.}
Table~\ref{tbl:topics} shows the estimated half-lives $\gamma_k$ and most representative words for five selected topics in the {\tt ml} dataset that are associated with at least 100 threads.  Fig.~\ref{fig:posttimes} plots the total number of posts made on these topics each week during the course. 

We observe topics with half-lives ranging from hours to weeks. We can use these timescales to categorize threads: course content-related topics (Topics~1 and 2) mostly have short half-lives of hours, small-talk topics (Topics 3 and 4) stay active for longer with half-lives of around one day, and course logistics topics (Topic~5) have much longer half-lives of around one week. Activities in threads on course content-related topics develop and decay rapidly, since they are most likely spurred by specific course materials or assignments. For example, posts on Topic~1 are about implementing gradient descent, which is covered in the second and third weeks of the course, and posts on Topic~2 are about neural networks, which is covered in the fourth and fifth weeks.  Small-talk discussions are extremely common at the beginning and the end of the course, while course logistics discussions (e.g., concerning technical issues) are less frequent but steady in volume throughout the course.

\paragraph{Excitation from notifications.}
Table~\ref{tbl:alphabeta} shows the estimated additional excitation induced by new activity notifications ($\widehat{\alpha}$) and explicit replies ($\widehat{\beta}$). In each course, we see that notifications increase the likelihood of participation significantly; for example, in \texttt{ml}, a learner's likelihood of posting after an explicit reply is 473 times higher than without any notification.  Notice also that $\widehat{\beta}$ is lowest while $\widehat{\alpha}$ is highest in \texttt{comp}. This observation is consistent with the fact that in humanities courses like \texttt{comp} the discussions in each thread will tend to be longer \cite{slnefficiency}, leading to more new activity notifications, while in engineering courses like \texttt{ml} and \texttt{algo} we would expect learners to more directly answer each other's questions, leading to more explicit replies.

}

{

\section{Conclusions and Future Work}
In this paper, we proposed a point processed-based probabilistic model for MOOC discussion forum posts, and demonstrated its performance in thread recommendation and analytics using real-world datasets. Possible avenues of future work include (i) jointly analyzing discussion forum data and time-varying learner grades \cite{tracekdd,sparfatop} to better quantify the ``flow of knowledge'' between learners, (ii) incorporating up-votes and down-votes on the posts into the model, and (iii) leveraging the course syllabus to better model the emergence of new threads.  }

\balance
\bibliographystyle{ACM-Reference-Format}
\bibliography{moocforum}

\end{document}